 \documentclass[preprint2,numberedappendix,twocolappendix]{aastex}

\usepackage{natbib}

\newcommand{\oiii}[1]{[{\ensuremath{\mathrm{O}}}\,\textsc{iii}]\:#1}
\newcommand{\oii}[1]{[{\ensuremath{\mathrm{O}}}\,\textsc{ii}]\:#1}

\newcommand{\sii}[1]{[{\ensuremath{\mathrm{S}}}\,\textsc{ii}]\:#1}

\newcommand{\heii}[1]{{\ensuremath{\mathrm{He}}}\,\textsc{ii}\:#1}

\newcommand{\hii}{{\ensuremath{\mathrm{H}}}\,\textsc{ii}}

\newcommand{\hb}{\ensuremath{\mathrm{H}\beta}}
\newcommand{\ha}{\ensuremath{\mathrm{H}\alpha}}

\newcommand{\civ}[1]{[{\ensuremath{\mathrm{C}}}\,\textsc{iv}]\:#1}
\newcommand{\nv}[1]{[{\ensuremath{\mathrm{N}}}\,\textsc{v}]\:#1}
\newcommand{\lya}{\ensuremath{\mathrm{Ly}\alpha}}

\shorttitle{}
\shortauthors{Shirazi et al.}

\begin{document}

\title{STARS WERE BORN IN SIGNIFICANTLY DENSER REGIONS IN THE EARLY UNIVERSE}

\author{Maryam Shirazi$^{1,2}$, Jarle Brinchmann$^1$, and Alireza Rahmati$^{1,3}$}
\affil{
$^1$ Leiden Observatory, Leiden University, P.O. Box 9513, 
2300 RA Leiden, The Netherlands;\\
$^2$ Department of Physics, Institute for Astronomy, ETH Zurich, CH-8093 Zurich, Switzerland;\\
$^3$ Max-Planck Institute for Astrophysics, Karl-Schwarzschild-Strasse 1, 85748 Garching, Germany;\\
Received 2013 July 17; accepted for publication in ApJ}
\email{shirazi@strw.leidenuniv.nl}

\begin{abstract}
The density of the warm ionized gas in high-redshift galaxies is known to be higher than what is typical in local galaxies on similar scales. At the same time, the mean global properties of the high- and low-redshift galaxies are quite different. Here, we present a detailed differential analysis of the ionization parameters of 14 star-forming galaxies at redshift 2.6--3.4, compiled from the literature. For each of those high-redshift galaxies, we construct a comparison sample of low-redshift galaxies closely matched in specific star formation rate (sSFR) and stellar mass, thus ensuring that their global physical conditions are similar to the high-redshift galaxy. We find that the median $\log \oiii 5007/ \oii 3727$ line ratio of the high-redshift galaxies is 0.5 dex higher than their local counterparts. We construct a new calibration between the $\oiii 5007/\oii 3727$ emission line ratio and ionization parameter to estimate the difference between the ionization parameters in the high and low-redshift samples. Using this, we show that the typical density of the warm ionized gas in star-forming regions decreases by a median factor of $7.1^{+10.2}_{-5.4}$ from $z\sim 3.3$ to $z\sim 0$ at fixed mass and sSFR. We show that metallicity differences cannot explain the observed density differences. Because the high- and low-redshift samples are comparable in size, we infer that the relationship between star formation rate density and gas density must have been significantly less efficient at $z\sim2-3$ than what is observed in nearby galaxies with similar levels of star formation activity.
\end{abstract}


\section{Introduction}

The cosmic star formation rate (SFR), averaged over all observed galaxies in the Universe, has dropped by a factor of $>10$ during the last $\sim$ 10 Gyr \citep[e.g.,][]{Hopkins-Beacon06}. In addition to the increasing fraction of actively star-forming galaxies with increasing look-back time, the SFRs of typical galaxies increases rapidly toward the earlier stages of galaxy formation \citep[e.g.,][]{Noeske07,Daddi07,Elbaz07,Elbaz11}. Several studies also provide hints that star formation conditions in distant galaxies (i.e., $z\sim 2-3$) are significantly different from the nearby Universe; emission lines from ionized gas in and around star-forming regions show different characteristics in distant and nearby galaxies \citep[e.g.,][]{Brinchmann08,Liu08,Newman13}, actively star-forming galaxies show higher gas fractions at higher redshifts \citep[e.g.,][]{Tacconi10,Genzel10}, and clumpy star-forming disks become increasingly more prevalent at higher redshifts \citep[e.g.,][]{Cowie95,Elmegreen06,Genzel11,Wisnioski12}. The average density of the warm ionized gas in typical high-redshift (high-$z$) galaxies is also known to be significantly higher than in typical low-redshift (low-$z$) galaxies on similar scales \citep[e.g.,][]{Elmegreen09, Lehnert09, Letiran11, Newman12, Tacconi13, Lehnert13}.

These studies have revealed that distant star-forming galaxies form a population of objects that are distinct from their nearby analogs. However, it is unclear whether the main difference between low-$z$ and high-$z$ star-forming galaxies is related to their strongly evolving global properties, such as stellar mass \citep[e.g.,][]{Ilbert13,Muzzin13}, SFR \citep[e.g.,][]{Noeske07,Daddi07,Elbaz07,Elbaz11} or metallicity \citep[e.g.,][]{Mannucci10,Lara-Lopez10}, or that the interstellar medium (ISM) conditions were significantly different in similar galaxies at high-$z$. Comparing representative samples of high-$z$ and low-$z$ star-forming galaxies \citep[e.g.][]{Rigby11} cannot disentangle the evolution in global characteristics from the possibly evolving star formation conditions. We address this issue by selecting a comparison ensemble of low-$z$ galaxies for each high-$z$ star-forming galaxy in our sample, ensuring that the stellar mass and star formation activities are similar in our high-$z$ galaxies and their low-$z$ comparison samples. This allows us to evaluate the differences in star formation conditions between the high-$z$ star-forming galaxies and their nearby analogs. 

Although observations of some lensed galaxies at high-$z$ reach spatial resolutions of $\sim$100 pc \citep[e.g.,][]{Swinbank09,Jones10,Wuyts14}, even this spatial resolution is insufficient to directly compare the small-scale properties of the ISM in high-$z$ and low-$z$ star-forming galaxies. However, these properties can be constrained through their impact on the emission line spectra of galaxies \citep[e.g.,][]{YehMatzner2012}. Here we use emission line ratios to derive the average ionization parameter of star-forming regions. Since the ionization parameter is a measure of ionizing radiation intensity per unit density, we can use it to constrain the density of star-forming regions in distant galaxies and compare it with that of their nearby counterparts. 

The structure of the paper is as follows. In Section \ref{sec:Data}, we introduce our high-$z$ sample and explain how we select their low-$z$ counterparts. In Section \ref{sec:Method}, we introduce our new calibration for calculating the ionization parameter using the emission line ratios. In Section \ref{sec:results}, we present our main results and compare the density of ionized gas in high-$z$ and nearby galaxies. In Section \ref{sec:metals}, we investigate the impact of metallicity variations between the high-$z$ and local galaxy samples on our results. In Section \ref{sec:discussion}, we discuss the implications of our finding and in Section \ref{sec:conclusions}, we end the paper with concluding remarks.

\section{Data}
\label{sec:Data}
We have assembled a sample of 14 high-$z$ star-forming galaxies from the literature for which published $\oii{\lambda3727}$, $\oiii{\lambda{5007}}$, and $\hb$ emission line fluxes are available (they have \break $\oiii{\lambda{5007}}/\hb>0$). This sample consists of two galaxies (RXJ1053 and, Cl0949) from \citet[][R11]{Richard11}; seven galaxies from the AMAZE sample \citep[][M08]{Maiolino08}; four galaxies from the LSD sample \cite[][M09]{Mannucci09}, and the 8 o'clock arc \break \citep[][\textit{arc}]{DZ11,Shirazi13}. These galaxies span redshifts between $z = 2.39$ and $z = 3.69$ with a median redshift of $z = 3.39$. All these galaxies also have gas metallicity, stellar mass and SFR estimates. To test our results further, we also use a sample of three galaxies in the SINS survey \citep{Forster09,Forster11} that have directly measured electron densities using the \textsc{[S\,ii]} doublet \citep{Lehnert09}. The physical properties of our high-$z$ sample are summarized in Table \ref{tab:highz}.

\begin{figure}[t]
\centerline{\hbox{\includegraphics[width=0.4\textwidth, angle=90] {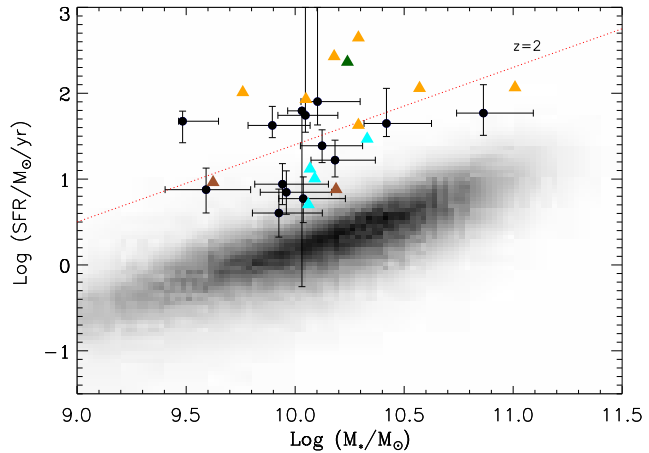}}}
\caption{High-$z$ sample (colored symbols; see the left panel in Figure \ref{hiz-lowz} for definition of different colours) and the median values of the low-$z$ samples (black circles) in the $\rm{M}_*-\rm{SFR}$ plane are shown where main-sequence star-forming SDSS galaxies are shown as a 2D histogram. Error bars shown on the black circles indicate 1 $\sigma$ scatter in the low-$z$ sample of each high-$z$ galaxy. The fit to the main-sequence $z\sim2$ star-forming galaxies \citep{Elbaz11} is shown by a dashed red line.}
\label{fig:mass-sfr}                     	
\end{figure}

We compare these galaxies to matched samples of low-$z$ galaxies from the Sloan Digital Sky Survey \citep[SDSS;][]{York00}. We used the MPA-JHU\footnote{http://www.mpa-garching.mpg.de/SDSS/DR7} value added catalogues \citep{Brinchmann04, Tremonti04} for SDSS DR7 \citep{Abazajian09} as our parent sample and selected star-forming galaxies following \citet{Brinchmann04}, with the adjustments of the line flux uncertainties given in \citet{Brinchmann13}. Furthermore, we used SDSS DR8 \citep{Aihara11} photometry to estimate stellar masses. The median and 1 $\sigma$ scatter of the physical properties of the low-$z$ sample of each high-$z$ galaxy are summarized in Table \ref{tab:lowz}. In Figure \ref{fig:mass-sfr} we show our high-$z$ and low-$z$ samples in the $\rm{M}_*-\rm{SFR}$ plane where main-sequence $z\sim0$ (SDSS) and the fit to the main-sequence $z\sim2$ star-forming galaxies \citep{Elbaz11} are shown as well.
	
As argued above, it is essential to take out correlations with global properties of galaxies when comparing their ISM conditions. To achieve this, we select for each high-$z$ galaxy, all star-forming galaxies in the SDSS DR7 that have $\log M_*$ and $\log (\mathrm{SFR}/M_*)$  within 0.3 dex of that of the high-$z$ galaxy. According to the results of the SFR-mass relation \citep[e.g.,][]{Noeske07, Zahid12}, for a given mass, $10^{10} \;M{\odot}$, the sSFR increases by $\sim0.5$ dex from $z \sim 0.1$ to $z \sim 0.8$ and by $\sim1.3$ dex from $z \sim 0.1$ to $z \sim2.2$. The scatter of the SFR-mass relation is about 0.3 dex. Therefore, by requiring the sSFR of the local and high-$z$ sample differs by no more than 0.3 dex, the two samples can be considered as having similar sSFR.

We also require that the SDSS galaxies have $z>0.02$ so that $\oii{\lambda{3727, 29}}$ are measured; they also have $\oiii{\lambda{5007}}/\hb>0$. By default, we do not explicitly constrain the low-$z$ samples to match the metallicity and/or size of their high-$z$ counterparts as this would reduce the size of our sample, and in the case of metallicity, it is subject to systematic uncertainties \citep[e.g.,][]{Kewley08}. However, as we show below, matching metallicities and/or sizes does not affect our results significantly.

Any significant contribution of ionizing radiation from an active galactic nucleus (AGN) could bias our estimates of the ionization parameter. For the low-$z$ sample we can exclude strong AGNs using the BPT diagram \citep{BPT}. At high-$z$, the galaxies from M08 and R11 do not show any evidence indicating the presence of AGNs in their rest frame UV spectra (i.e., $\nv$,$\civ$,$\heii$ or broad $\lya$), X-ray, and 24 $\mu m$ $Spitzer$-MIPS observations \citep{Maiolino08, Richard11, Shirazi13}. The LSD galaxies also show no evidence of AGN activity in X-ray observations \citep{Mannucci09}. While the aforementioned arguments do not rule out the presence of some AGN activity that is optically thick for X-rays, this is unlikely to significantly influence the optical emission lines which originate in only moderately obscured regions. One galaxy from the SINS sample (Q2343-BX610) that we use in this study has an indication of possible AGNs from mid-IR observations \citep{Forster11,Hainline12} and from an analysis of resolved spectroscopy presented by \cite{Newman13}. However, we note that we do not use our calibration to infer ionization parameter for the SINS galaxies. Thus, we conclude that AGN activity is unlikely to bias our results at high-$z$.

Since we match samples where the physical parameters have been inferred using different authors, it is also important that these differences do not lead to significant biases. Note that we are not directly comparing high-$z$ galaxies with each other, thus our focus is on comparisons between high and low redshift. However, we note that as most of the galaxies in our high-$z$ sample (11 out of 14) are selected from the AMAZE/LSD surveys, the physical parameters for these are already on the same scale.

The stellar masses in \citet{Maiolino08} were derived using a Salpeter initial mass function (IMF) and were adjusted to a Chabrier IMF \citep{Chabrier03} in agreement with that used for the SDSS galaxies and in \citet{Mannucci09}. The stellar masses were also all derived using spectral energy distribution (SED) fitting and \citet{BC03} models, but the methodology used differed between the low- and high-$z$ galaxies. For the high-$z$ galaxies, the models used for SED fitting use smooth star formation histories and the best fit model was chosen to infer physical parameters. For the low-z sample, in contrast, a library of star formation histories consisting of smooth histories with superposed bursts from \citet{Gallazzi05, Gallazzi08} was used, and physical parameters were inferred using a Bayesian approach. 

For the stellar masses, \citet{Pozzetti07} have argued that star formation histories incorporating bursts result in slightly higher masses (mean of 0.17 dex) than best fit models using smooth star formation histories. We do not correct for this here but note that doing so would lead us to select slightly higher-mass low-$z$ galaxies and would slightly strengthen our results. We note that \citet{Pacifici12} find a similar effect but with the opposite sign when Bayesian analysis is used in both cases. For our present needs, however, the argument in \citet{Pozzetti07} is the relevant one.

The SFRs in \citet{Mannucci09} were inferred from emission lines, but they show that there is good agreement between SFRs derived from emission lines and those calculated based on SED fitting for AMAZE/LSD galaxies (see their Figure 3). For the SDSS, a similar result was found by \citet{Salim07} for star-forming galaxies, as is relevant for us, and we use emission line determined SFRs here.

\begin{figure*}
\centerline{\hbox{\includegraphics[trim= 1cm 0cm 1cm 5cm, clip=clip,width=0.7\textwidth, angle=0] {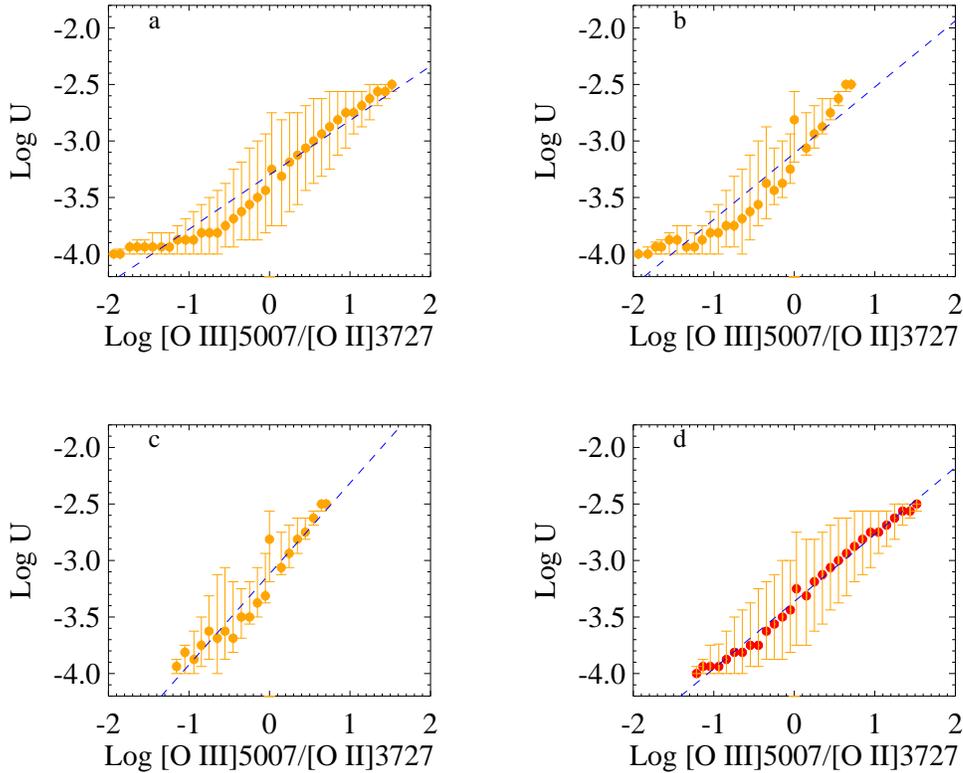}}}
             
	\caption{Best fit relations between ionization parameters and the 
	$\oiii{\lambda{5007}}/\oii{\lambda3727}$ (O32) ratios are shown by blue dashed lines. The top left panel shows 
	the best fit using all CL01 models ($0.01 < \tau_{V} < 4$), the 
         top right panel shows the best fit using only models with $\tau_{V} \sim 0.2$,
         the bottom left panel shows the best fit using only models with $\tau_{V} \sim 0.2$ 
         and $\xi \sim 0.3$ (i.e., Galactic dust-to-metal ratio), and the
         bottom right panel shows the best fit to all models with $\xi \sim 0.3$. 
         The results in the paper are presented for the fit shown in the bottom right panel.}
\label{fig:logu-o3o2}                     	
\end{figure*}

\section{Methodology}
\label{sec:Method}
The high-$z$ galaxies all have measured \break $\oiii{\lambda{5007}}$ and $\oii{\lambda3727}$ line fluxes. This allows us to use the strong sensitivity of the $\oiii{\lambda{5007}}/\oii{\lambda3727}$ (hereafter O32) ratio to the ionization parameter \citep{Penston90} to estimate this. \citet{Kewley02} derived an estimator for the ionization parameter using the dereddened O32 ratio. Since this can not easily be applied to our high-$z$ sample in the absence of reliable reddening estimates, here we calibrate a new relation between the ionization parameter and the observed O32 ratio using the \citet[][hereafter CL01]{CL01} models that account for variations in dust properties and metallicities (see Table 4 in \cite{Shirazi12} for the CL01 model grid used in our study). CL01 use the photoionisation code Cloudy \citep{Ferland98} and construct their models by varying effective parameters that describe the ensemble of \hii\ regions and the diffuse ionized gas in a galaxy. These effective parameters depend on time due to time-dependent stellar ionizing radiation. The effective ionization parameter, defined as the ratio of the hydrogen-ionizing photon production rate to the gas density,  in these models is taken to be the volume-averaged ionization parameter over the Str{\"o}mgren sphere (see Equation 9 and 10 in CL01).

We wish to construct a calibration between the ionization parameter and O32 ratio that treats the metallicity as a free parameter. Based on this approach, as long as our high-$z$ and low-$z$ samples do not differ greatly in metallicity, we do not need to know this exactly. We discuss this assumption further below, but given that, we still have several possible ways to construct the calibration from the CL01 models.

{\textit{a-}} Leaving all parameters in the CL01 models as free parameters in the fitting procedure (including all dust attenuation parameters, $0.01 < \tau_{V} < 4$ ). This is likely to give a large amount of scatter in the relationship.

{\textit{b-}} Using only models with $\tau_{V} \sim 0.2$ and leaving all other model parameters free. This fit is appropriate if line ratios are corrected for dust attenuation but there is no constraint on the dust-to-metal ratio ($\xi$).

 {\textit{c-}} Using only models with $\tau_{V} \sim 0.2$ and $\xi \sim 0.3$ (i.e., the Galactic 
dust-to-metal ratio) and leaving all other model parameters free. Since $\xi$ is expected to evolve weakly with time \citep{Calura08}, it is reasonable to fix its value. 

{\textit{d-}} Using $\xi \sim 0.3$ and leaving all other model parameters free. Since $\xi$ is likely not to differ strongly from this value, this is the best choice for a calibration when the amount of dust attenuation is unknown. 

These fits are plotted in Figure \ref{fig:logu-o3o2} from the top left to the bottom right, respectively. The best fits for the relation between ionization parameter and $\rm{Log} \;\oiii{\lambda{5007}}/\oii{\lambda3727}$ (Log O32) are summarized as Equations (1) to (4), respectively. We use option {\it{d}}, Equation (\ref{eq:Ueq}), as our reference in this study because, in general, we do not have enough information to accurately constrain the dust attenuation for the high-$z$ galaxy sample. Quantitatively, varying the dust attenuation from $\tau_{V}=0.5$ to $\tau_{V}=1.55$ will lead to a difference (reduction) of $\sim0.15$ dex in the $\rm{Log \;U}$ at fixed Log O32.

To derive our reference relation we fix $\xi=0.3$, which is the Galactic value \citep[see][for a discussion]{Brinchmann13}, and allow all other parameters to vary. We use the same fit for estimating the ionization parameter for low-$z$ counterparts of high-$z$ galaxies.
\begin{equation}
     \rm{Log \;U} = -3.300 \pm 0.017 + (0.481 \pm 0.019) \; Log \;O32
     \label{eq:Ueq1}
     \end{equation}
\begin{equation}
     \rm{Log \;U} = -3.109 \pm 0.039 + (0.586 \pm 0.039) \; Log \;O32
     \label{eq:Ueq2}
     \end{equation}
  \begin{equation}
     \rm{Log \;U} = -3.119 \pm 0.027 + (0.804 \pm 0.035) \; Log \;O32
     \label{eq:Ueq3}
     \end{equation}
  \begin{equation}
     \rm{Log \;U} = -3.363 \pm 0.011 + (0.593 \pm 0.012) \;Log \;O32
     \label{eq:Ueq}
  \end{equation}

This assumes that the average $\tau_{V}$ is the same in the low- and high-$z$ samples. This is not an entirely unreasonable assumption given the very similar physical properties of the samples, but it does warrant further attention. To do this we have compared the $A_V$ values derived from SED fitting to the high-$z$ galaxies to the $A_V$ inferred from both SED fitting of the low-$z$ counter parts and the Balmer decrement of for these. The average difference $A_V(\mbox{high}-z)-A_V(\mbox{low}-z)$ is $0.28\pm 0.23$  when $A_V$ from SED fitting is used at low-$z$ and is $0.22 \pm 0.23$ when $A_V$ from the Balmer decrement is used and scaled down by a factor of 0.44 \citep{Calzetti97} to account for the difference in attenuation of emission lines and continuum. These differences in dust attenuation lead to changes in $\log U$ of $\sim 0.04$ dex, which is smaller than the differences we see below, so we have chosen to not attempt to correct for this since it is not clear how the SED fit $A_V$ values at high-$z$ should be converted to emission line attenuation corrections.

We are primarily focused on relative statements in this work so the most important aspect of these calibrations is how they convert relative statements in O32 to relative statements about $\log U$. Since the slope in Equations (\ref{eq:Ueq1}), (\ref{eq:Ueq2}), and (\ref{eq:Ueq}) is similar, they will result in similar relative statements about $\log U$, while that in Equation (\ref{eq:Ueq3}) is even steeper and would lead to an even stronger result than that outlined below.

	\begin{figure*}
\centerline{\hbox{\includegraphics[trim= 1cm 0cm 10cm 2cm, clip=clip,width=0.3\textwidth, angle=90] {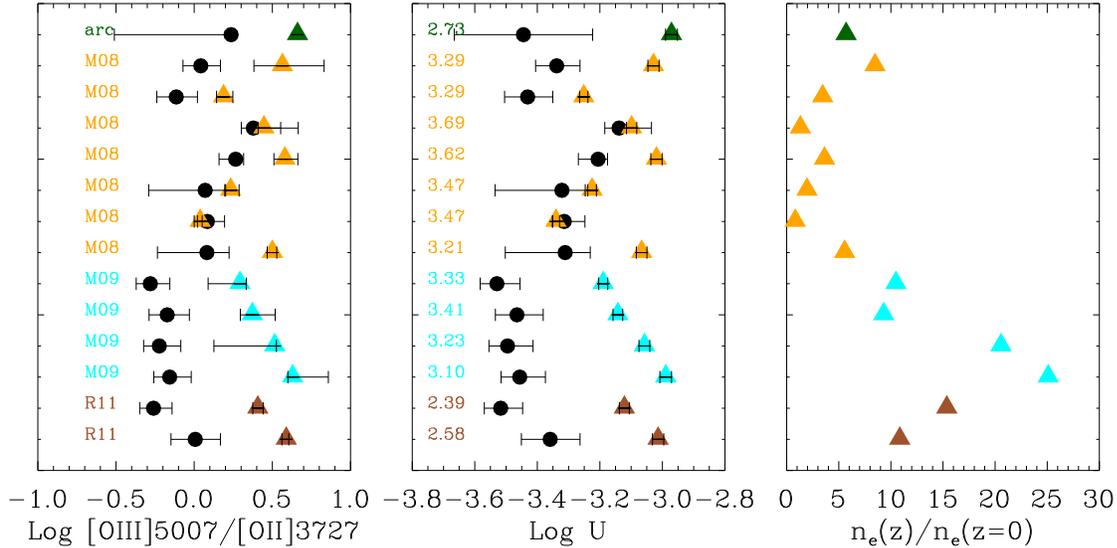}}}
         
	 \caption{Comparison between $\oiii{\lambda{5007}}/\oii{\lambda3727}$ ratio, ionization parameter, and electron density at low-$z$ and high-$z$. The $x$ axis of the left panel shows the $\oiii{\lambda{5007}}/\oii{\lambda3727}$ ratio, the middle panel shows the ionization parameter, and the right panel shows the electron density at high-$z$ relative to that of low-$z$. Colored symbols show high-$z$ galaxies with their redshift indicated and black circles show the median values for the low-$z$ sample of each high-$z$ galaxy. Error bars span from the 16\% to the 84\% confidence level. We see that high-$z$ galaxies show higher $\oiii{\lambda{5007}}/\oii{\lambda3727}$ ratios than their low-$z$ analogs (up to $\approx 0.8$ dex higher), even though their masses and sSFR are the same. The middle panel shows the ionization parameters derived using our new calibration between the $\oiii{\lambda{5007}}/\oii{\lambda3727}$ ratio and the ionization parameter. We see that high-$z$ galaxies show up to $\sim$ 0.5 dex higher (median $\sim$ 0.3 dex) ionization parameters than their low-$z$ analogs. This translates to up to 25 times higher electron density for high-$z$ galaxies relative to their low-$z$ analogs. }
 	\label{hiz-lowz}                     	
	\end{figure*}

	  \begin{figure*}
\centerline{\hbox{\includegraphics[trim= 0cm 0cm 8cm 2cm, clip=clip,width=0.5\textwidth, angle=90] {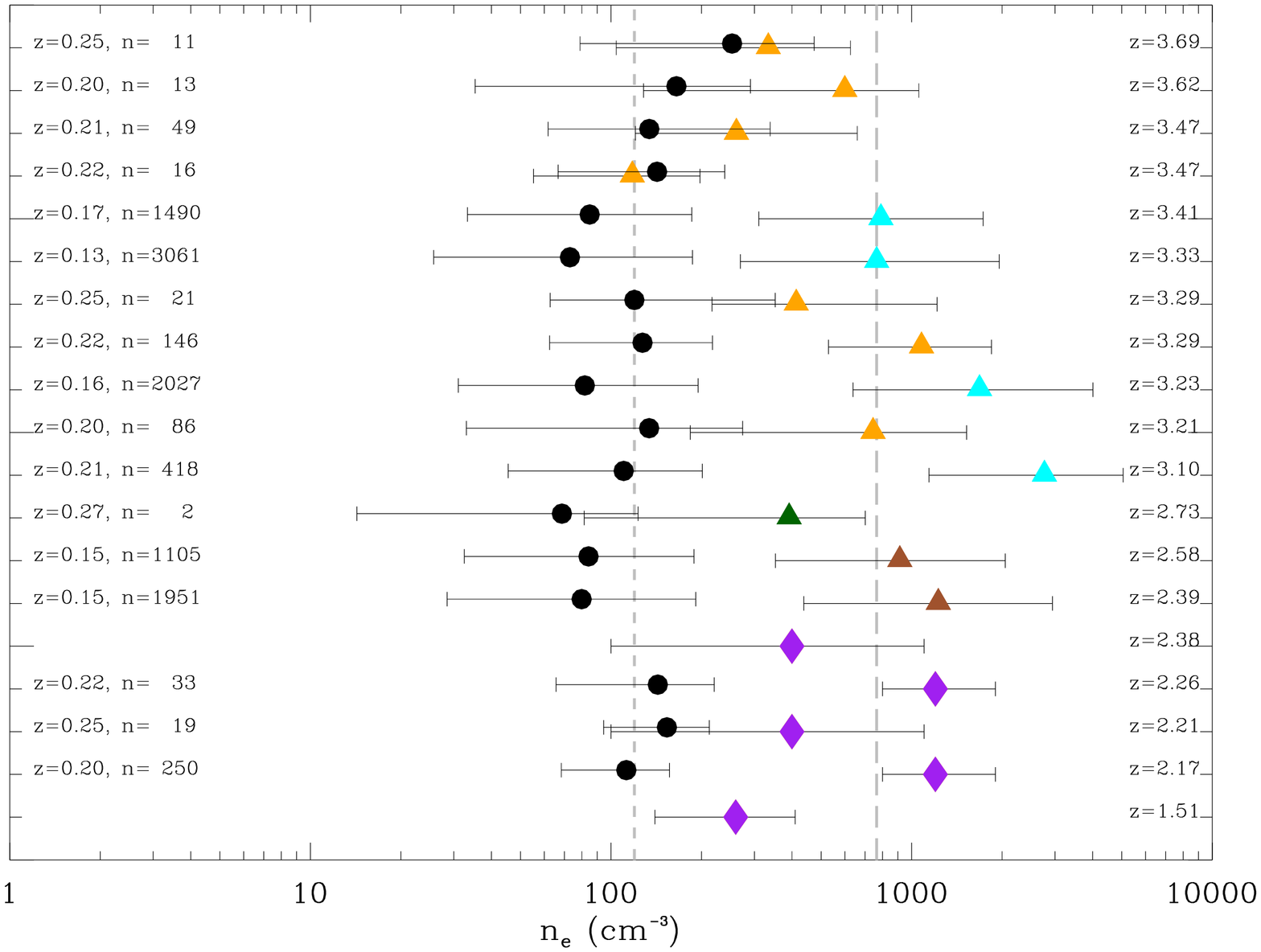}}}
         
          \caption{Median value of the electron density for the low-$z$ samples inferred from the $\sii{6716,6731}$ doublet is shown by the black filled circles. The high-$z$ values are inferred from the low-$z$ values multiplying by $n_{e}(z)/n_e(z=0)$ ratios shown in Figure~\ref{hiz-lowz}. Colored symbols show our high-$z$ sample sorted based on their redshifts from bottom to top as indicated in the figure. Five galaxies from the SINS survey that have directly measured electron densities are shown by purple diamonds. The median values of the redshifts of low-$z$ samples are shown in black and the number of low-$z$ analogs in each sample are indicated with $n$. Error bars span from the 16\% to the 84\% confidence level (low-$z$ data; they show scatter in the sample, high-$z$ data; they show propagation of uncertainties through Equation \ref{eq:Ueq}). The grey small-dashed and long-dashed lines show the median value for the electron density at low-$z$ and high-$z$, respectively.}
	\label{electron-density}                     	
	\end{figure*}

\section{Results}
\label{sec:results}

The left and middle panels of Figure~\ref{hiz-lowz} compare the $\oiii{\lambda{5007}}/\oii{\lambda3727}$ ratios and corresponding ionization parameters (from Equation~\ref{eq:Ueq}) for our high-$z$ sample (colored symbols), and the median values of their low-$z$ analogs (black circles). Error bars shown on the black circles indicate 1 $\sigma$ scatter in the low-$z$ sample of each high-$z$ galaxy. It is evident that the high-$z$ star-forming galaxies show significantly higher $\oiii{\lambda{5007}}/\oii{\lambda3727}$ ratios (up to $\approx 0.8$ dex higher) compared to their low-$z$ analogs. This translates into significantly higher ionization parameters (up to $\sim$ 0.5 dex) in the high-$z$ galaxies relative to low-$z$ galaxies even though their SFRs and masses are constrained to be the same.

For a given production rate of hydrogen ionizing photons, $Q$, and after assuming that most of the ionizing photons are absorbed locally, the ionization parameter in a typical ionized region can be related to the hydrogen number density, $n_{\rm{H}}$,
\begin{equation}
\rm{U}^3 \propto Q(t) \;n_{\rm{H}} \; \epsilon^{2},
\label{eq:UQ}
\end{equation}
where $\epsilon$ is the volume filling factor of the ionized gas, which is defined as the ratio between the volume-weighted and mass-weighted average hydrogen densities \citep{CL01}. This allows us to constrain the densities of star-forming regions, by measuring their ionization parameters.

Assuming that the production rate of hydrogen ionizing photons and volume filling factors of the ionized gas are similar in typical star-forming regions in high-$z$ galaxies and their low-$z$ analogs, one can translate the ratio between the ionization parameters of the high-$z$ galaxies and their low-$z$ counterparts into the ratio of their ionized gas densities. The difference between the density of the ionized gas in star-forming regions in our high-$z$ galaxies and their low-$z$ analogs is shown in the right panel of Figure \ref{hiz-lowz}. This shows up to $\approx$ 25 times higher densities in high-$z$ star-forming galaxies. 

To derive physical densities for our high-$z$ galaxies from the relative density differences shown in Figure \ref{hiz-lowz}, we exploit the fact that for the nearby galaxies we can estimate the electron density from the $\sii{\lambda{6716,6731}}$ ratio and thus get an estimate of the electron density in the high-$z$ galaxies. The resulting absolute densities for the ionized gas in our high-$z$ star-forming galaxies are shown in Figure~\ref{electron-density}. The median values of the electron densities of the low-$z$ samples, inferred from the $\sii{6716,6731}$ doublet, are shown by the black filled circles in the figure where error bars show 1 $\sigma$ scatter. The median values of the redshifts of the low-$z$ samples and the number of low-$z$ analogs in each sample are indicated with $n$ in the figure. Colored symbols show our high-$z$ sample with their redshifts indicated. The high-$z$ values are inferred from the low-$z$ values multiplied by $n_{e}(z)/n_e(z=0)$ ratios shown in Figure~\ref{hiz-lowz}, and their error bars show propagation of uncertainties based on Equation \ref{eq:Ueq}. The grey small-dashed and long-dashed lines show the median values for the electron density at low-$z$ and high-$z$, respectively.  

Besides the sensitivity of the ionization parameter to the density of the emitting gas, it also depends on the production rate of ionizing photons and the volume filling factor of the ionized gas \citep{CL01,Kewley13a, Kewley13b}. Therefore, our density estimates might also be sensitive to the possible differences in the ionizing photons production rate and the volume filling factor of the ionized gas between high-$z$ and nearby galaxies. In particular, the geometry of the gas distribution can affect the volume filling factor which is largely unconstrained even at low redshift. However, there is not a particular reason to have a strong redshift-dependent volume filling factor in systems with similar SFRs and stellar masses. In addition to the volume filling factor, the SED of the ionizing radiation could change the amount of $\rm{O}^+$ ionizing photons and hence change our results. Noting that the SED is metallicity dependent and the metallicities of our high and low redshift samples are approximately the same (see Section \ref{sec:metals}) we expect the SED of the ionizing radiation also to be similar in our high- and low-$z$ samples. Nonetheless, there are remaining questions such as whether the \hii\ regions are radiation or density bounded \citep{Nakajima12} which we cannot claim are controlled by the way we select our samples.

To address the above mentioned concerns from a different angle, in Figure~\ref{electron-density}, we show electron densities for a sample of five high-$z$ star-forming galaxies in the SINS survey \citep{Forster09,Lehnert09} as purple diamonds. The electron density for these galaxies has been measured directly using the $\sii{\lambda{6716,6731}}$ doublet and is in a good agreement with our inferred evolution in density estimated from the ionization parameter. For three of these five objects that have available stellar masses and sSFRs \citep{Forster11}, we constructed low-$z$ analog samples. The comparison between the electron density of these three objects and their low-$z$ analogs also shows good agreement (evolution in density with a median factor of 8.4) with the density ratios we obtained for our high-$z$ star-forming galaxies using their ionization parameters (an evolution in density with a median factor of 7.1). This further strengthens our argument that an elevated density of star-forming regions in high-$z$ galaxies is the main reason for their higher ionization parameter.

\section{Metallicity dependence}
\label{sec:metals}

  A key result in this work is that high-$z$ galaxies typically have a 0.5 dex higher Log O32 than low-$z$ galaxies with the same mass and sSFR. We interpret this as primarily being due to a difference in ionization parameter, but O32 is also sensitive to metallicity. Ideally, we would select our high-$z$ and low-$z$ samples to have the same metallicity, but to do this, we require a metallicity estimator that can be applied equally at low-$z$ and high-$z$ allowing for a variation in ionization parameter. With the current data available for high-$z$ galaxies, this is not possible; thus we need to assess whether metallicity differences between the samples could be the reason for the observed offset.  
 
 \citet{Mannucci10} and \citet{Lara-Lopez10} showed that there is a relationship between stellar mass, metallicity and SFR that appear to hold to high-$z$ ($z<2.5$ for Mannucci et al. and $z<3.5$ for Lara-Lopez et al.). Therefore, if this holds for our galaxies, a selection on stellar mass and SFR should ensure that the metallicity difference between the high- and low-$z$ sample is small.  Given our small sample and considering that \citet{Mannucci10} argued that the multi-parameter relationship was not well established at $z>2.5$, where most of our high-$z$ galaxies lie, it is necessary to examine this assumption more carefully. It is useful to start this by asking what metallicity difference would give a O32 difference similar to what is observed. From \citet[][their Figure 8]{Brinchmann08}, or directly using the CL01 models, we find that a change in metallicity from $1~Z_{\odot}$ to $0.1~Z_{\odot}$ leads to a change in Log O32 of $0.40\pm0.07$ dex. Thus, we need a major difference in metallicity to explain the results.

We can test for a large offset in metal content by calculating the metallicities of the high- and low-$z$ samples in a consistent way. To do this we adopt the methodology used for AMAZE and LSD described in \citep{Maiolino08} for both high- and low-$z$ galaxies. Note that, by construction, this method assumes the same relation between O32 and metallicity at low-$z$ and high-$z$. Therefore, by using it, we will maximize the contribution of metallicity to the change in O32 and hence derive a minimum difference in ionization parameter between the low- and high-$z$ objects. Based on the derived metallicities, we can calculate the maximum difference in O32 between high- and low-$z$ galaxies due to metallicity differences, using the CL01 models and by averaging over $U$. This gives us the expected change in O32 due to metallicity only, and we subtract this off the actual observed difference for each galaxy.
 	\begin{figure}[t]
\centerline{\hbox{\includegraphics[width=0.5\textwidth, angle=0] {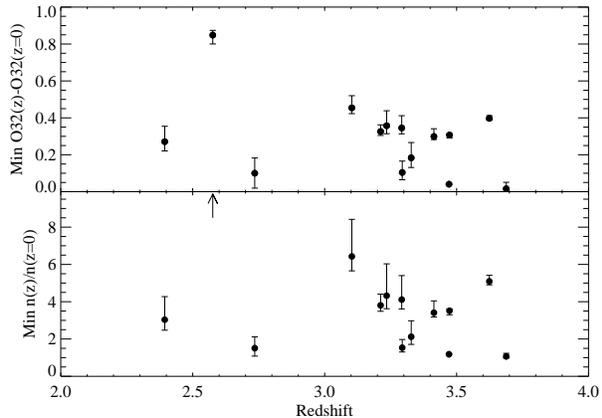}}}
	\caption{Top panel: the minimum difference in O32 between the high- and low-$z$ samples when corrected for metallicity as described in the text. Bottom panel: the resulting minimum density difference between the high- and low-$z$ samples when corrected for metallicity.}
	\label{fig:do32-min}                     	
	\end{figure}

The resulting difference can be seen in the top panel of  Figure~\ref{fig:do32-min}. We emphasize that since we have used an abundance calibration that assumes that changes in O32 are due to metallicity, this correction should be the maximum possible correction. This gives a lower limit to the difference in O32 between high- and low-$z$ galaxies, and it is still quite sizeable. Converting this to a density difference as done in the main text, we get the bottom panel in that figure. This shows that the mean (median) electron density of the high-$z$ galaxies is 5.5 (3.5) times higher than the low-$z$ galaxies with the same sSFR and mass.

To further test the sensitivity of our results to metallicity differences between our high-$z$ galaxies and their low-$z$ analogs, we made a low-$z$ comparison sample for all high-$z$ galaxies, ensuring that their metallicities were equal to within 0.3 dex, in addition to matching their stellar masses and sSFRs\footnote{Note that this additional metallicity constraint decreases the low-$z$ sample sizes.}. In this case, we found that high-$z$ galaxies show a median of $\approx 6.1$ higher density compared to their low-$z$ analogs with similar sSFRs, masses, and metallicities; a result which is not significantly different from what we found without matching metallicities.

We also note that the densities that are measured directly from the \textsc{[S\,ii]} doublet for the five high-$z$ galaxies we selected from the SINS, are not derived using our calibration and hence are insensitive to variations in metallicity. Yet they have densities which are on average 8.4 times higher than their local analogs. It also worth noting that not all SINS galaxies have detected \textsc{[S\,ii]} which is consistent with these conclusions because $\textsc{[S\,ii]}/\ha$ decreases with increasing $U$ at fixed metallicity \citep[e.g.,][their Figure 11]{B08WR}.

In conclusion, regardless of how we correct for possible differences in metallicity between the high- and low-$z$ samples, the effect is minor and the main result of the paper is robust to these corrections. Thus, we conclude that differences in metallicity can not explain the observed major offset in O32, and that systematic differences in the ionization parameter is the main cause.

\begin{figure}[t]
\centerline{\hbox{\includegraphics[width=0.4\textwidth, angle=90] {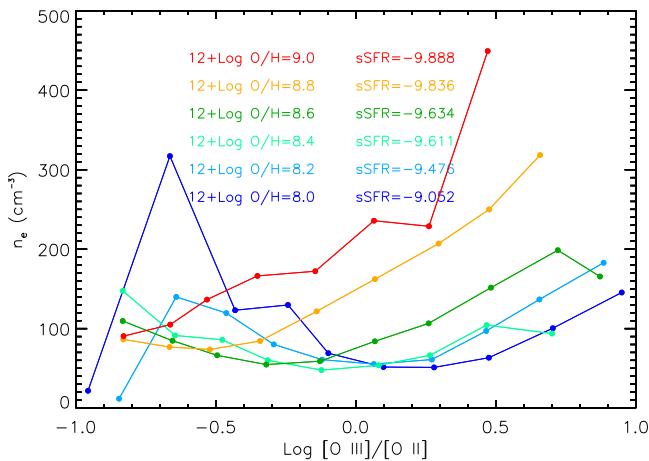}}}
	\caption{Correlation between the metallicity and gas density and O32 in star-forming galaxies in the SDSS that have measured gas density using the $\sii$ doublet is shown. We bin the data in gas-phase metallicity ($12+\log \rm{O/H}=8$ to 9) as derived in \citet{Tremonti04}. The figure illustrates quite clearly that at fixed metallicity, an increase in O32 corresponds to an increase in the electron density. }
	\label{fig:high_ne_bin}                     	
	\end{figure}

We also study the correlation between the metallicity and gas density and O32 in star-forming galaxies in the SDSS that have measured gas density using the $\sii$ doublet. This is shown in Figure \ref{fig:high_ne_bin}. We bin the data in gas-phase metallicity ($12+\log \rm{O/H}=8$ to 9) measured as derived in \citet{Tremonti04}. The figure illustrates that at fixed metallicity, an increase in O32 corresponds to an increase in the electron density \footnote{We note that the highest electron density allowed in the CL01 models is $n_e=100 \;\rm{cm^{-3}}$.}. 

\section{Discussion}
\label{sec:discussion}

The observed strong evolution in the global properties such as star formation intensity, stellar mass and size indicates that mean star formation conditions are different in distant galaxies compared to typical galaxies today \citep{Cowie95,Elmegreen06,Noeske07,Daddi07,Elbaz07,Elbaz11,Tacconi10,Genzel10,Genzel11}. In this work, however, we have shown that even when the star formation intensity and mass are the \emph{same}, the density in the ionized gas in high- and low-$z$ galaxies differ dramatically. This difference would naturally imply a higher pressure in the colder ISM surrounding the ionized gas \citep{Dopita06}, and thence its higher density. 

This could naturally occur if star formation at high-$z$ was more concentrated to the central regions, so to check this we compared the $u$-band half-light radius  for the SDSS galaxies with the half-light radius of the high-$z$ galaxies when they are available (for seven galaxies). Among the high-$z$ galaxies, only one has a smaller size than the median size of its low-$z$ counterparts. This is in agreement with the findings of \citet{Lehnert09} and cannot explain the density differences seen for any reasonable mass profile in the galaxies. We double-checked this by constructing matched low-$z$ samples that have $\log \mathrm{SFR}/\pi r_{1/2}^2$ within 0.3 dex of their high-$z$ counterparts, where $r_{1/2}$ is the half-light radius. This results in a median density difference greater than 19 between low-$z$ and high-$z$ galaxies compared to a median difference of $\sim 7$ before matching SFR densities. This shows that size differences are unlikely to be the explanation of the systematic differences. We have not required a match in SFR density in the bulk of the paper. However, because the size definitions are somewhat arbitrary, and we do not have sizes for all galaxies at high-$z$.

Assuming now that the distribution of star formation is comparable at low- and high-$z$, we next assume that the \textsc{H\,ii} regions are in pressure equilibrium with their surrounding ISM \citep{OandC97,Dopita06}. Under this assumption the increased pressure in the ionized regions implies a higher pressure in the cold ISM. There are considerable uncertainties in how ionized regions expand in detail. However, in our case it is not unreasonable to assume that those complexities should be similar at high and low redshift. This is because the evolution of the \textsc{H\,ii} regions is driven by the energy injection from massive stars which should be similar at high-$z$ and low-$z$, given how we selected our samples. The same applies to cosmic ray production rates which contribute to the heating (and the pressure support) of the ambient ISM. Note that this also means that the contribution of radiation pressure to the equilibrium for the \textsc{H\,ii} regions \citep[e.g.][]{YehMatzner2012} should be similar at low and high redshift.

It is hard to test whether the \textsc{H\,ii} regions in the high-$z$ galaxies have reached pressure equilibrium with their surrounding ISM. 
However, since the mechanical input energy is the same at high and low redshift, and the life-times of the relevant stellar population is also the same, it seems unlikely that the evolutionary age of the \textsc{H\,ii} regions differs significantly between the high-$z$ and low-$z$ samples. This is also supported by  \citet{Verdolini13}, who used a population study to show that the line emission of a galaxy will typically be dominated by the youngest  \textsc{H\,ii} regions. \citet{Verdolini13} also show explicitly the effect of an elevated ambient pressure on emission line ratios (their Figure 8), which is a qualitatively similar trend to what we infer here.

Thus, the simplest explanation for the elevated density in the high-$z$ \textsc{H\,ii} regions is an elevated pressure in the cold ISM relative to similar galaxies nearby. This increased pressure could arise from various sources, but in general, one would expect a pressure-density relation, $P\propto \rho^\gamma$ with $\gamma > 1$. Thus, the increased pressure would correspond to an increased ISM density by an amount that depends on the model adopted for the ISM, and we do not attempt to discuss this in detail here. The simplest model, where the ISM temperature is the same at high and low redshift, would predict that the density difference between the ISM at high- and low-$z$ would be the same as that of the the ionized regions, i.e., $\rho_{\mathrm{high}-z} \sim 7 ~\rho_{\mathrm{low}-z}$. 

This conclusion has important implications for empirical star formation law as well. The most popular scaling relation observed between star formation activity and gas surface density in the local Universe is the Kennicutt--Schmidt relation \citep{Kennicutt98},
\begin{equation}
  \Sigma_{\mathrm{SFR}} \propto \Sigma_{\mathrm{gas}}^{1.4},
\end{equation}
where $\Sigma$ denote surface densities. In our case $\Sigma_{\mathrm{SFR}}$ is approximately the same in the high- and low-$z$ galaxies (see above), but the gas density is much higher. If the scale-height of the gas is not significantly smaller in high-$z$ galaxies, one can conclude that the scaling relation in the high-$z$ galaxies is significantly different from what is observed in their low-$z$ counterparts, being a factor $\sim 5$--$7$ less efficient.

We note, however, that we can not distinguish between molecular and atomic gas. Therefore, our results are for the total gas and we cannot directly compare them to molecular studies \citep[e.g.,][]{Daddi10,Tacconi13} at high-$z$ and leave a discussion of this for future work.

\section{Conclusion}
\label{sec:conclusions}

In this work we compare the physical conditions of the ISM in high-$z$ galaxies and their low-$z$ counterparts that are selected to have similar global properties as that of high-$z$ galaxies. This selection criteria minimize the differences between distant and nearby galaxies due to the evolution of the global properties such as mass and sSFR from high-$z$ to low-$z$ and can therefore be used to study the evolution of intrinsic properties of the ISM.  

Previous studies have  already pointed out that the physical densities/properties of the star-forming regions at high-$z$ are very different from those in the local Universe and we confirm this here. Using a novel approach, we have been able to go one step further, and show that this difference can not fully be explained by an increased star formation activity in the high-$z$ galaxies. Since we compare high and low-$z$ galaxies that are matched in sSFR, their different densities must reflect an intrinsic difference in ISM conditions between high and low-$z$. We argue that this difference is primarily due to a difference in the density of the warm ionized gas. We have also shown that the differences between the high- and low-$z$ galaxies can not be explained by differences in metallicity.  By showing that the high-$z$ and low-$z$ samples are also comparable in size, we conclude that the relationship between SFR density and gas density must have been significantly less efficient at $z\sim2-3$ than what is observed locally. This, in turn, implies that most of the stars in the local Universe were formed following a different star formation scaling relation than what is observed in normal galaxies today.

\section*{Acknowledgments}
We thank the anonymous referees for very insightful comments which improved this paper. We thank Marijn Franx for helpful discussion. We express our appreciation to Leslie Sage for his help and suggestions.

\begin{deluxetable}{llllllllllll}
\tabletypesize{\scriptsize}
\rotate
\tablecaption{High-$z$ Sample.\label{tab:highz}}
\tablewidth{0pt}
\tablehead{
ID & Name & z & Log Mass & Log sSFR & SFR & $\Sigma_{\rm{SFR}}$ & $r_{1/2}$ & Log O32 & $\rm{12+Log O/H}$ & Log U & $n_{e}$ \\
 &  &  & ($M_{\odot}$) & ($\rm{yr}^{-1}$) & ($M_{\odot}\;\rm{yr}^{-1}$) & ($M_{\odot}\;\rm{yr}^{-1}\;\rm{kpc}^{-2}$) & ($\rm{kpc}$) & & & & ($\rm{cm}^{-3}$)}
 \startdata
R11 & RXJ1053         & 2.576 & $ 9.62_{ -0.72}^{  0.75}$ & $-8.67$ & $  9.1_{  -2.3}^{  2.3}$ & 0.22 & 3.62$\pm$ 0.45 & $0.589_{-0.0}^{  0.0}$ & $8.68_{ -0.12}^{  0.11}$ & $-3.01_{ -0.02}^{  0.02}$ & $ 913.8_{  -561.1}^{  1135.8}$\\
R11 & Cl0949          & 2.394 & $10.19_{ -0.18}^{  0.22}$ & $-9.31$ & $  7.5_{  -1.5}^{  1.5}$ & 0.19 & 3.50$\pm$ 0.88 & $0.407_{-0.0}^{  0.0}$ & $8.10_{ -0.05}^{  0.06}$ & $-3.12_{ -0.02}^{  0.02}$ & $1226.8_{  -789.0}^{  1714.7}$\\
M09 & SSA22a-C30      & 3.103 & $10.33_{ -0.38}^{  0.31}$ &$ -8.87 $& $ 29.0_{ -21.0}^{ 81.0}$ & 4.21 & 1.48$\pm$ 0.44 & $0.630_{-0.0}^{  0.2}$ & $8.16_{ -0.60}^{  0.20}$ & $-2.99_{ -0.02}^{  0.02}$ & $2766.6_{ -1623.8}^{  2286.8}$\\
M09 & Q0302-C131      & 3.235 & $10.09_{ -0.33}^{  0.10}$ & $-9.09$ & $ 10.0_{  -4.0}^{  6.0}$ & 1.97 & 1.27$\pm$ 0.37 & $0.515_{-0.4}^{  0.0}$ & $8.00_{ -0.40}^{  0.25}$ & $-3.06_{ -0.02}^{  0.02}$ & $1682.9_{ -1044.2}^{  2325.3}$\\
M09 & Q0302-M80       & 3.414 & $10.07_{ -0.19}^{  0.23}$ &$ -8.95$ & $ 13.0_{  -8.0}^{ 17.0}$ & 7.36 & 0.75$\pm$ 0.24 & $0.372_{-0.1}^{  0.1}$ & $8.36_{ -0.15}^{  0.15}$ & $-3.14_{ -0.02}^{  0.02}$ & $ 790.5_{  -480.5}^{   938.7}$\\
M09 & Q0302-C171      & 3.328 & $10.06_{ -0.28}^{  0.10}$ & $-9.36 $& $  5.0_{  -2.0}^{  2.0}$ & 1.02 & 1.25$\pm$ 0.39 & $0.293_{-0.2}^{  0.0}$ & $8.14_{ -0.45}^{  0.25}$ & $-3.19_{ -0.01}^{  0.01}$ & $ 765.2_{  -495.7}^{  1192.1}$\\
M08 & CDFa-C9         & 3.212 & $ 9.95_{ -0.08}^{  0.40}$ & $-7.53$ & $265.0_{   0.0}^{  0.0}$ &   \nodata & \nodata& $0.500_{-0.0}^{  0.0}$ & $8.10_{ -0.21}^{  0.18}$ & $-3.07_{ -0.02}^{  0.02}$ & $ 744.5_{  -560.9}^{   779.0}$\\
M08 & CDFS-4414       & 3.471 & $10.34_{ -0.22}^{  0.19}$ &$ -8.29 $& $113.0_{   0.0}^{  0.0}$ &   \nodata & \nodata & $0.038_{-0.0}^{  0.0}$ & $8.54_{ -0.14}^{  0.15}$ & $-3.34_{ -0.01}^{  0.01}$ & $ 117.8_{   -62.6}^{    80.0}$\\
M08 & CDFS-4417       & 3.473 & $10.06_{ -0.11}^{  0.37}$ & $-7.42 $& $438.0_{   0.0}^{  0.0}$ &   \nodata & \nodata & $0.233_{-0.0}^{  0.1}$ & $8.55_{ -0.10}^{  0.09}$ & $-3.22_{ -0.01}^{  0.01}$ & $ 261.3_{  -140.9}^{   398.4}$\\
M08 & CDFS-16767      & 3.624 & $ 9.82_{ -0.16}^{  0.10}$ & $-7.89 $& $ 84.0_{   0.0}^{  0.0}$ &   \nodata & \nodata & $0.580_{-0.1}^{  0.1}$ & $8.31_{ -0.17}^{  0.11}$ & $-3.02_{ -0.02}^{  0.02}$ & $ 599.5_{  -471.2}^{   457.3}$\\
M08 & CDFS-2528       & 3.688 & $ 9.53_{ -0.07}^{  0.09}$ & $-7.52$ & $101.0_{   0.0}^{  0.0}$ &   \nodata & \nodata & $0.446_{-0.0}^{  0.2}$ & $8.07_{ -0.28}^{  0.39}$ & $-3.10_{ -0.02}^{  0.02}$ & $ 333.7_{  -229.6}^{   292.2}$\\
M08 & SSA22a-M38      & 3.294 & $10.78_{ -0.41}^{  0.18}$ & $-8.72$ & $115.0_{   0.0}^{  0.0}$ &   \nodata & \nodata & $0.188_{-0.0}^{  0.1}$ & $8.34_{ -0.12}^{  0.15}$ & $-3.25_{ -0.01}^{  0.01}$ & $ 413.3_{  -196.4}^{   801.8}$\\
M08 & SSA22a-aug16M16 & 3.292 & $10.06_{ -0.21}^{  0.20}$ & $-8.44 $& $ 42.0_{   0.0}^{  0.0}$ &   \nodata & \nodata & $0.564_{-0.2}^{  0.3}$ & $7.99_{ -0.34}^{  0.26}$ & $-3.03_{ -0.02}^{  0.02}$ & $1078.7_{  -549.2}^{   765.0}$\\
arc & 8oclock         & 2.735 & $10.24_{  0.45}^{ -1.80}$ & $-7.88 $& $228.0_{ -10.0}^{ 10.0}$ & 9.26 & 2.80$\pm$ 0.20 & $0.661_{}^{ }$ & $8.35_{ -0.19}^{  0.19}$ & $-2.97_{ -0.02}^{  0.02}$ & $ 391.5_{  -310.0}^{   310.0}$\\
SINS\tablenotemark{a} & Q2343-BX389 & 2.172 & $10.61_{ -2.16}^{  0.77}$ & $-9.22$ & \nodata & \nodata & \nodata & \nodata & \nodata & \nodata & $1200.0_{ -400}^{  700}$\\
SINS\tablenotemark{a} & Q2343-BX610 & 2.210 & $11.00_{ -0.60}^{  2.70}$ & $-9.22$ & \nodata & \nodata & \nodata & \nodata & \nodata & \nodata & $ 400.0_{ -300}^{  700}$\\
SINS\tablenotemark{a} & Q2346-BX482 & 2.256 & $10.26_{ -0.46}^{  0.79}$ & $-8.36$ & \nodata & \nodata & \nodata & \nodata & \nodata & \nodata & $1200.0_{ -400}^{  700}$\\
\enddata
\tablenotetext{a}{We use the properties of the SINS galaxies that are given in \cite{Forster11} and \cite{ Lehnert09}.}
\tablenotetext{b}{Stellar masses from \cite{Maiolino08} are scaled to \citet{Chabrier03} IMF.}

\end{deluxetable}

 \begin{deluxetable}{llllllllllll}
\tabletypesize{\scriptsize}
\rotate
\tablecaption{Low-$z$ Sample.\label{tab:lowz}}
\tablewidth{0pt}
\tablehead{
 High-$z$ ID & $\langle$ z $\rangle$\tablenotemark{a} & $\langle$Log Mass$\rangle$ & $\langle$Log sSFR$\rangle$ & $\langle$Log SFR$\rangle$ & $\langle$$\Sigma_{SFR}$$\rangle$ & $\langle$$r50_{u}$$\rangle$ & $\langle$Log O32$\rangle$ &  $\langle$$\rm{12+Log O/H}$$\rangle$& $\langle$Log U$\rangle$ & $\langle$$n_{e}$$\rangle$ \\
  &  & ($M_{\odot}$) & ($\rm{yr}^{-1}$) & ($M_{\odot}\;\rm{yr}^{-1}$) & ($M_{\odot}\;\rm{yr}^{-1}\;\rm{kpc}^{-2}$) & ($\rm{kpc})$ & & M08 calib \tablenotemark{b}& & ($\rm{cm}^{-3}$) \\}
 \startdata
 
 RXJ1053         & $0.15_{-0.07}^{ 0.07}$ & $ 9.59_{ -0.19}^{  0.20}$ & $-8.81$ & $  0.9_{  -0.3}^{  0.2}$ &  1.89 & $1.12_{-0.24}^{ 0.41}$ & $ 0.007_{-0.155}^{ 0.161}$ & $8.49_{-0.05}^{ 0.12}$ & $-3.36_{ -0.09}^{  0.10}$ & $  84.2_{  -51.7}^{  104.7}$\\
Cl0949          & $0.15_{-0.05}^{ 0.07}$ & $10.04_{ -0.11}^{  0.20}$ & $-9.32$ & $  0.8_{  -0.3}^{  0.3}$ &  0.93 & $1.46_{-0.42}^{ 0.65}$ & $-0.260_{-0.088}^{ 0.118}$ & $8.70_{-0.07}^{ 0.05}$ & $-3.52_{ -0.05}^{  0.07}$ & $  79.9_{  -51.4}^{  111.6}$\\
SSA22a-C30      & $0.21_{-0.08}^{ 0.05}$ & $10.18_{ -0.11}^{  0.19}$ & $-9.03$ & $  1.2_{  -0.2}^{  0.2}$ &  3.70 & $1.22_{-0.27}^{ 0.32}$ & $-0.157_{-0.101}^{ 0.138}$ & $8.65_{-0.10}^{ 0.07}$ & $-3.46_{ -0.06}^{  0.08}$ & $ 110.2_{  -64.7}^{   91.1}$\\
Q0302-C131      & $0.16_{-0.05}^{ 0.07}$ & $ 9.96_{ -0.12}^{  0.21}$ & $-9.19 $& $  0.8_{  -0.3}^{  0.2}$ &  1.35 & $1.32_{-0.33}^{ 0.54}$ & $-0.223_{-0.100}^{ 0.137}$ & $8.67_{-0.09}^{ 0.07}$ & $-3.50_{ -0.06}^{  0.08}$ & $  81.8_{  -50.8}^{  113.1}$\\
Q0302-M80       & $0.17_{-0.06}^{ 0.06}$ & $ 9.94_{ -0.13}^{  0.21}$ & $-9.08$ & $  0.9_{  -0.2}^{  0.2}$ &  1.87 & $1.24_{-0.29}^{ 0.40}$ & $-0.173_{-0.116}^{ 0.142}$ & $8.64_{-0.11}^{ 0.08}$ & $-3.47_{ -0.07}^{  0.08}$ & $  84.9_{  -51.6}^{  100.9}$\\
Q0302-C171      & $0.13_{-0.05}^{ 0.07}$ & $ 9.93_{ -0.12}^{  0.20}$ & $-9.37 $& $  0.6_{  -0.3}^{  0.3}$ &  0.53 & $1.53_{-0.46}^{ 0.84}$ & $-0.281_{-0.090}^{ 0.125}$ & $8.71_{-0.08}^{ 0.05}$ & $-3.53_{ -0.05}^{  0.07}$ & $  73.0_{  -47.3}^{  113.8}$\\
CDFa-C9         & $0.20_{-0.06}^{ 0.05}$ & $ 9.80_{ -0.11}^{  0.22}$ & $-8.38$ & $  1.6_{  -0.2}^{  2.3}$ & 12.93 & $0.91_{-0.91}^{ 0.25}$ & $ 0.081_{-0.316}^{ 0.142}$ & $8.44_{-8.44}^{ 0.04}$ & $-3.31_{ -0.19}^{  0.08}$ & $ 133.9_{ -100.9}^{  140.2}$\\
CDFS-4414       & $0.22_{-0.12}^{ 0.07}$ & $10.16_{ -0.11}^{  0.21}$ & $-8.47$ & $  1.7_{  -0.2}^{  0.2}$ & 13.82 & $1.01_{-0.10}^{ 0.90}$ & $ 0.084_{-0.064}^{ 0.110}$ & $8.48_{-0.04}^{ 0.11}$ & $-3.31_{ -0.04}^{  0.07}$ & $ 142.3_{  -75.6}^{   96.7}$\\
CDFS-4417       & $0.21_{-0.07}^{ 0.06}$ & $ 9.90_{ -0.09}^{  0.20}$ & $-8.39$ & $  1.6_{  -0.2}^{  2.2}$ & 14.91 & $0.96_{-0.96}^{ 0.25}$ & $ 0.070_{-0.361}^{ 0.126}$ & $8.44_{-8.44}^{ 0.06}$ & $-3.32_{ -0.21}^{  0.07}$ & $ 134.1_{  -72.3}^{  204.4}$\\
CDFS-16767      & $0.20_{-0.09}^{ 0.06}$ & $ 9.67_{ -0.08}^{  0.22}$ & $-8.08 $& $  1.7_{  -0.2}^{  0.2}$ & 17.04 & $1.00_{-0.24}^{ 0.21}$ & $ 0.265_{-0.106}^{ 0.051}$ & $8.44_{-0.09}^{ 0.02}$ & $-3.21_{ -0.06}^{  0.03}$ & $ 165.1_{ -129.8}^{  125.9}$\\
CDFS-2528       & $0.25_{-0.03}^{ 0.03}$ & $ 9.37_{ -0.08}^{  0.09}$ &$ -7.71$ & $  1.7_{  -0.0}^{  0.1}$ & 29.83 & $0.81_{-0.20}^{ 0.28}$ & $ 0.378_{-0.077}^{ 0.175}$ & $8.34_{-0.24}^{ 0.10}$ & $-3.14_{ -0.05}^{  0.10}$ & $ 252.9_{ -174.0}^{  221.4}$\\
SSA22a-M38      & $0.25_{-0.01}^{ 0.04}$ & $10.62_{ -0.13}^{  0.15}$ & $-8.88$ & $  1.7_{  -0.2}^{  0.4}$ & 10.05 & $1.13_{-0.20}^{ 0.69}$ & $-0.115_{-0.124}^{ 0.135}$ & $8.65_{-0.07}^{ 0.06}$ & $-3.43_{ -0.07}^{  0.08}$ & $ 119.6_{  -56.9}^{  232.1}$\\
SSA22a-aug16M16 & $0.22_{-0.06}^{ 0.05}$ & $ 9.90_{ -0.09}^{  0.18}$ &$ -8.62$ & $  1.4_{  -0.2}^{  0.2}$ &  7.30 & $1.09_{-0.22}^{ 0.20}$ & $ 0.042_{-0.114}^{ 0.125}$ & $8.49_{-0.05}^{ 0.10}$ & $-3.34_{ -0.07}^{  0.07}$ & $ 127.3_{  -64.8}^{   90.3}$\\
8oclock         & $0.27_{-0.22}^{ 0.00}$ & $10.03_{ -0.07}^{  0.00}$ & $-8.04$ & $  1.8_{  -2.0}^{  0.0}$ & 17.63 & $3.01_{-1.95}^{ 0.00}$ & $ 0.236_{-0.747}^{ 0.000}$ & $8.78_{-0.40}^{ 0.00}$ & $-3.44_{ -0.22}^{  0.22}$ & $  68.7_{  -54.4}^{   54.4}$\\
 Q2343-BX389 & $0.20_{-0.08}^{ 0.05}$ & $10.44_{ -0.10}^{  0.17}$ &$ -9.27$ & $  1.2_{  -0.2}^{  0.2}$ &  2.49 & $1.40_{-0.36}^{ 0.51}$ & $-0.205_{-0.085}^{ 0.113}$ & $8.70_{-0.08}^{ 0.04}$ & $-3.48_{ -0.05}^{  0.07}$ & $ 112.5_{ -68}^{ 127}$\\
    Q2343-BX610 & $0.25_{-0.09}^{ 0.02}$ & $10.84_{ -0.11}^{  0.27}$ & $-9.27$ & $  1.5_{  -0.1}^{  0.2}$ &  5.23 & $1.55_{-0.52}^{ 0.65}$ & $-0.197_{-0.061}^{ 0.140}$ & $8.71_{-0.09}^{ 0.04}$ & $-3.48_{ -0.04}^{  0.08}$ & $ 153.5_{ -94}^{ 228}$\\
    Q2346-BX482 & $0.22_{-0.10}^{ 0.06}$ & $10.07_{ -0.06}^{  0.13}$ & $-8.55$ & $  1.6_{  -0.1}^{  0.1}$ & 12.15 & $1.08_{-0.20}^{ 0.27}$ & $ 0.084_{-0.087}^{ 0.110}$ & $8.47_{-0.03}^{ 0.09}$ & $-3.31_{ -0.05}^{  0.07}$ & $ 143.1_{ -65}^{ 100}$\\
\enddata

\tablenotetext{a}{$\langle$  $\rangle$ shows the median value where lower and upper values show $1-\sigma$ scatter in the sample.}
\tablenotetext{b}{Metallicities are measured using \cite{Maiolino08} calibration.}
\end{deluxetable}


\clearpage
\begin{thebibliography}{}
\expandafter\ifx\csname natexlab\endcsname\relax\def\natexlab#1{#1}\fi

\bibitem[{Abazajian} {et~al.}(2009)]{Abazajian09}
{Abazajian}, K.~N., {Adelman-McCarthy}, J.~K., {Ag{\"u}eros}, M.~A., {et~al.}
  2009, \apjs, 182, 543

\bibitem[{Aihara} {et~al.}(2011)]{Aihara11}
{Aihara}, H., {Allende Prieto}, C., {An}, D., {et~al.} 2011, \apjs, 193, 29

\bibitem[{Baldwin, Phillips \& Terlevich}(1981)]{BPT}
Baldwin, J. A., Phillips, M. M., \& Terlevich, R. 1981, PASP, 93, 5
   
\bibitem[{Brinchmann} {et~al.}(2013)]{Brinchmann13}
{Brinchmann}, J., {Charlot}, S., {Kauffmann}, G., {et~al.} 2013, \mnras, 432, 2112

\bibitem[{{Brinchmann} {et~al.}(2004)}]{Brinchmann04}
{Brinchmann}, J., {Charlot}, S., {White}, S.~D.~M., {et~al.} 2004, \mnras, 351,
  1151

\bibitem[{{Brinchmann} {et~al.}(2008{\natexlab{a}}){Brinchmann}, {Kunth}, \&
  {Durret}}]{B08WR}
{Brinchmann}, J., {Kunth}, D., \& {Durret}, F. 2008{\natexlab{a}}, \aap, 485,
  657

\bibitem[{{Brinchmann} {et~al.}(2008{\natexlab{b}}){Brinchmann}, {Pettini}, \&
  {Charlot}}]{Brinchmann08}
{Brinchmann}, J., {Pettini}, M., \& {Charlot}, S. 2008{\natexlab{b}}, \mnras,
  385, 769

\bibitem[Bruzual \& Charlot(2003)]{BC03}
Bruzual, G., \& Charlot, S. 2003, MNRAS, 344, 1000

\bibitem[Chabrier(2003)]{Chabrier03} Chabrier, G.\ 2003, ApJL, 586, L133 

\bibitem[{{Calura} {et~al.}(2008)}]{Calura08}
{Calura}, F. and {Pipino}, A. and {Matteucci}, F., 2008, \aap, 479, 669

\bibitem[Calzetti et al.(1997)]{Calzetti97} Calzetti, D., Meurer, 
G.~R., Bohlin, R.~C., et al.\ 1997, \aj, 114, 1834 

\bibitem[{{Charlot} \& {Longhetti}(2001)}]{CL01}
{Charlot}, S., \& {Longhetti}, M. 2001, \mnras, 323, 887

\bibitem[{{Cowie} {et~al.}(1995){Cowie}, {Hu}, \& {Songaila}}]{Cowie95}
{Cowie}, L.~L., {Hu}, E.~M., \& {Songaila}, A. 1995, \aj, 110, 1576

\bibitem[{{Daddi} {et~al.}(2007)}]{Daddi07}
{Daddi}, E., {Dickinson}, M., {Morrison}, G., {et~al.} 2007, \apj, 670, 156

\bibitem[Daddi et al.(2010)]{Daddi10} Daddi, E., Elbaz, D., 
Walter, F., et al.\ 2010, ApJL, 714, L118 


\bibitem[{{Dessauges-Zavadsky} {et~al.}(2011)}]{DZ11}
{Dessauges-Zavadsky}, M., {Christensen}, L., {D'Odorico}, S., {Schaerer}, D.,
  \& {Richard}, J. 2011, \aap, 533, A15

\bibitem[{{Dopita} {et~al.}(2006)}]{Dopita06}
{Dopita}, M.~A., {Fischera}, J., {Sutherland}, R.~S., {et~al.} 2006, \apj, 647,
  244

\bibitem[{{Elbaz} {et~al.}(2007)}]{Elbaz07}
{Elbaz}, D., {Daddi}, E., {Le Borgne}, D., {et~al.} 2007, \aap, 468, 33

\bibitem[Elbaz et al.(2011)]{Elbaz11} 
Elbaz, D., Dickinson, M., Hwang, H.~S., et al.\ 2011, \aap, 533, A119 

\bibitem[{{Elmegreen} \& {Elmegreen}(2006)}]{Elmegreen06}
{Elmegreen}, B.~G., \& {Elmegreen}, D.~M. 2006, \apj, 650, 644

\bibitem[{{Elmegreen} {et~al.}(2009)}]{Elmegreen09}
Elmegreen, D.~M., 
Elmegreen, B.~G., Marcus, M.~T., et al.\ 2009, \apj, 701, 306 

\bibitem[\protect\citeauthoryear{Ferland et al.}{1998}]{Ferland98}
   Ferland, G. J., Korista, K. T., et al., 1998, PASP, 110, 761
   

\bibitem[{{F{\"o}rster Schreiber} {et~al.}(2009)}]{Forster09}
{F{\"o}rster Schreiber}, N.~M., {Genzel}, R., {Bouch{\'e}}, N., {et~al.} 2009,
  \apj, 706, 1364

\bibitem[{{F{\"o}rster Schreiber} {et~al.}(2011){F{\"o}rster Schreiber},
  {Shapley}, {Genzel}, {Bouch{\'e}}, {Cresci}, {Davies}, {Erb}, {Genel},
  {Lutz}, {Newman}, {Shapiro}, {Steidel}, {Sternberg}, \&
  {Tacconi}}]{Forster11}
{F{\"o}rster Schreiber}, N.~M., {Shapley}, A.~E., {Genzel}, R., {et~al.} 2011,
  \apj, 739, 45

\bibitem[Gallazzi et al.(2005)]{Gallazzi05}
  Gallazzi, A., Charlot, S., et al., 2005, MNRAS, 362, 41
%
 \bibitem[Gallazzi et al.(2008)]{Gallazzi08}
  Gallazzi, A., Brinchmann, J., et al., 2008, MNRAS, 383, 1439
  
\bibitem[{{Genzel} {et~al.}(2010)}]{Genzel10}
{Genzel}, R., {Tacconi}, L.~J., {Gracia-Carpio}, J., {et~al.} 2010, \mnras,
  407, 2091

\bibitem[{{Genzel} {et~al.}(2011)}]{Genzel11}
{Genzel}, R., {Newman}, S., {Jones}, T., {et~al.} 2011, \apj, 733, 101

\bibitem[{{Hainline} {et~al.}(2012){Hainline}, {Shapley}, {Greene}, {Steidel},
  {Reddy}, \& {Erb}}]{Hainline12}
{Hainline}, K.~N., {Shapley}, A.~E., {Greene}, J.~E., {Steidel}, C.~C.,
  {Reddy}, N.~A., \& {Erb}, D.~K. 2012, \apj, 760, 74
  
\bibitem[{{Hopkins} \& {Beacom}(2006)}]{Hopkins-Beacon06}
{Hopkins}, A.~M., \& {Beacom}, J.~F. 2006, \apj, 651, 142

\bibitem[Ilbert et al.(2013)]{Ilbert13} Ilbert, O., McCracken, 
H.~J., Le Fevre, O., et al.\ 2013, A\&A, 556, A55

\bibitem[{{Jones} {et~al.}(2010)}]{Jones10}
{Jones}, T.~A., {Swinbank}, A.~M., {Ellis}, R.~S., {Richard}, J., \& {Stark},
  D.~P. 2010, \mnras, 404, 1247

\bibitem[{{Kennicutt}(1998)}]{Kennicutt98}
{Kennicutt}, R.~C., Jr., 1998, \apj, 498, 541

\bibitem[{{Kewley} \& {Dopita}(2002)}]{Kewley02}
{Kewley}, L.~J., \& {Dopita}, M.~A. 2002, \apjs, 142, 35

\bibitem[{{Kewley} \& {Ellison}(2008)}]{Kewley08}
{Kewley}, L.~J., \& {Ellison}, S.~L. 2008, \apj, 681, 1183

\bibitem[Kewley et al.(2013a)]{Kewley13a} Kewley, L.~J., Dopita, 
M.~A., Leitherer, C., et al.\ 2013a, ApJ, 774,
100 

\bibitem[Kewley et al.(2013b)]{Kewley13b} Kewley, L.~J., Maier, 
C., Yabe, K., et al.\ 2013b, ApJL, 774, L10

\bibitem[{{Lara-L{\'o}pez} {et~al.}(2010)}]{Lara-Lopez10}
{Lara-L{\'o}pez}, M.~A., {Cepa}, J., {Bongiovanni}, A., {et~al.} 2010, \aap,
  521, L53

\bibitem[{{Lehnert} {et~al.}(2009)}]{Lehnert09}
{Lehnert}, M.~D., {Nesvadba}, N.~P.~H., {Le Tiran}, L., {et~al.} 2009, \apj, 699, 1660
  
\bibitem[Lehnert et al.(2013)]{Lehnert13} 
Lehnert, M.~D., Le Tiran, L., Nesvadba, N.~P.~H., et al.\ 2013, \aap, 555, A72 


\bibitem[Le Tiran et al.(2011)]{Letiran11} 
Le Tiran, L., Lehnert, M.~D., van Driel, W., Nesvadba, N.~P.~H., \& Di Matteo, P.\ 2011, A \& A, 534, L4

\bibitem[{{Liu} {et~al.}(2008)}]{Liu08}
{Liu}, X., {Shapley}, A.~E., {Coil}, A.~L., {Brinchmann}, J., \& {Ma}, C.-P.
  2008, \apj, 678, 758

\bibitem[{{Maiolino} {et~al.}(2008)}]{Maiolino08}
{Maiolino}, R., {Nagao}, T., {Grazian}, A., {et~al.} 2008, \aap, 488, 463

\bibitem[{{Mannucci} {et~al.}(2010)}]{Mannucci10}
{Mannucci}, F., {Cresci}, G., {Maiolino}, R., {Marconi}, A., \& {Gnerucci}, A.
  2010, \mnras, 408, 2115

\bibitem[{{Mannucci} {et~al.}(2009)}]{Mannucci09}
{Mannucci}, F., {Cresci}, G., {Maiolino}, R., {et~al.} 2009, \mnras, 398, 1915

\bibitem[Muzzin et al.(2013)]{Muzzin13} Muzzin, A., Marchesini, 
D., Stefanon, M., et al.\ 2013, ApJ, 777, 18

\bibitem[Newman et al.(2012)]{Newman12} Newman, S.~F., Shapiro 
Griffin, K., Genzel, R., et al.\ 2012, \apj, 752, 111 

\bibitem[Newman et al.(2013)]{Newman13} 
Newman, S.~F., Buschkamp, P., Genzel, R., et al.\ 2014, ApJ, 781, 21



\bibitem[{{Noeske} {et~al.}(2007)}]{Noeske07}
{Noeske}, K.~G., {Weiner}, B.~J., {Faber}, S.~M., {et~al.} 2007, ApJL, 660,
  L43

\bibitem[Nakajima et al.(2012)]{Nakajima12} Nakajima, K., Ouchi, 
M., Shimasaku, K., et al.\ 2012, \apj, 745, 12

\bibitem[{{Oey} \& {Clarke}(1997)}]{OandC97}
{Oey}, M.~S., \& {Clarke}, C.~J. 1997, \mnras, 289, 570

\bibitem[{Pacifici et al.(2012)}]{Pacifici12} 
Pacifici, C., Charlot, S., Blaizot, J., \& Brinchmann, J.\ 2012, \mnras, 421, 2002

\bibitem[{{Penston} {et~al.}(1990)}]{Penston90}
{Penston}, M.~V., {Robinson}, A., {Alloin}, D., {et~al.} 1990, \aap, 236, 53

\bibitem[Pozzetti et al.(2007)]{Pozzetti07} 
Pozzetti, L., Bolzonella, M., Lamareille, F., et al.\ 2007, \aap, 474, 443

\bibitem[{{Richard} {et~al.}(2011)}]{Richard11}
{Richard}, J., {Jones}, T., {Ellis}, R., {et~al.} 2011, \mnras, 413, 643

\bibitem[{{Rigby} {et~al.}(2011)}]{Rigby11}
{Rigby}, J.~R., {Wuyts}, E., {Gladders}, M.~D., {Sharon}, K., \& {Becker},
  G.~D. 2011, \apj, 732, 59

\bibitem[Salim et al.(2007)]{Salim07} Salim, S., Rich, R.~M., 
Charlot, S., et al.\ 2007, \apjs, 173, 267 

\bibitem[{{Shirazi} \& {Brinchmann}(2012)}]{Shirazi12}
{Shirazi}, M., \& {Brinchmann}, J. 2012, \mnras, 421, 1043

\bibitem[{{Shirazi} {et~al.}(2014)}]{Shirazi13}
{Shirazi}, M., {Vegetti}, S., {Nesvadba}, N., {et~al.} 2014, \mnras, 440, 2201

\bibitem[{{Swinbank} {et~al.}(2009)}]{Swinbank09}
{Swinbank}, A.~M., {Webb}, T.~M., {Richard}, J., {et~al.} 2009, \mnras, 400,
  1121

\bibitem[{{Tacconi} {et~al.}(2010)}]{Tacconi10}
{Tacconi}, L.~J., {Genzel}, R., {Neri}, R., {et~al.} 2010, \nat, 463, 781

\bibitem[{{Tacconi} {et~al.}(2013)}]{Tacconi13} 
Tacconi, L.~J., Neri, R., Genzel, R., et al.\ 2013, \apj, 768, 74 

\bibitem[{{Tremonti} {et~al.}(2004)}]{Tremonti04}
{Tremonti}, C.~A., {Heckman}, T.~M., {Kauffmann}, G., {et~al.} 2004, \apj, 613,
  898

\bibitem[{{Verdolini} {et~al.}(2013)}]{Verdolini13}
{Verdolini}, S., {Yeh}, S.~C.~C., {Krumholz}, M.~R., {Matzner}, C.~D. \& {Tielens}, A.~G.~G.~M., 2013, \apj, 769, 12
	
\bibitem[{{Wisnioski} {et~al.}(2012)}]{Wisnioski12} 
Wisnioski, E., Glazebrook, K., Blake, C., et al.\ 2012, \mnras, 422, 3339 

\bibitem[Wuyts et al.(2014)]{Wuyts14} Wuyts, E., Rigby, J.~R., Gladders, M.~D., \& Sharon, K.\ 2014, \apj, 781, 61 

\bibitem[Zahid et al.(2012)]{Zahid12} Zahid, H.~J., Bresolin, 
F., Kewley, L.~J., Coil, A.~L., \& Dav{\'e}, R.\ 2012, \apj, 750, 120 


\bibitem[Yeh 
\& Matzner(2012)]{YehMatzner2012} Yeh, S.~C.~C., \& Matzner, C.~D.\ 2012, \apj, 757, 108 


\bibitem[{{York} {et~al.}(2000)}]{York00}
{York}, D.~G., {Adelman}, J., {Anderson}, J.~E., Jr., {et~al.} 2000, \aj, 120,
  1579

\end{thebibliography}
\end{document}